\documentclass[,fleqn,usenatbib]{aa}  
\usepackage{lastpage}
\usepackage{widetext}
\usepackage{graphicx}
\usepackage{dcolumn}
\usepackage{bm}
\usepackage{natbib}
\usepackage{multirow}
\usepackage{url}
\usepackage{color}
\usepackage{xcolor}
\usepackage{}
\usepackage{float}
\usepackage{cancel}
\usepackage{amssymb}
\usepackage{amsmath}
\usepackage{colortbl} 
\usepackage{comment}
\usepackage{subfig}
\usepackage{mathptmx}
\usepackage{txfonts}
\usepackage[T1]{fontenc}
\usepackage{ae,aecompl}
\usepackage{amsfonts}
\usepackage{academicons}
\usepackage{hyperref}
\usepackage{xcolor}
\usepackage{graphicx}
\usepackage[pdf]{pstricks}
\usepackage{epstopdf}
\usepackage{amssymb}
\usepackage{latexsym}

\usepackage{graphicx}
\usepackage{txfonts}
\newcommand{\vq}{\vec{q}}

\begin{document}

   \title{The cosmic web from perturbation theory}

   \author{F.-S. Kitaura
    \inst{1}\inst{2}\thanks{fkitaura@iac.es}
          \and
          F. Sinigaglia\inst{1}\inst{2}\inst{3}\inst{4}
          \and
          A. Balaguera-Antolínez
          \inst{1}\inst{2}
\and
G. Favole
           \inst{1}\inst{2}
          }

   \institute{Instituto de Astrof\'{\i}sica de Canarias, s/n, E-38205, La Laguna, Tenerife, Spain \email{fkitaura@iac.es} \and
 Departamento de Astrof\'{\i}sica, Universidad de La Laguna, E-38206, La Laguna, Tenerife, Spain
 \and 
Department of Physics and Astronomy, Università degli Studi di Padova, Vicolo dell’Osservatorio 3, I-35122, Padova, Italy 
 \and
INAF—Osservatorio Astronomico di Padova, Vicolo dell’Osservatorio 5, I-35122, Padova, Italy
             }

   \date{Received XX XX, 20XX; accepted XX XX, 20XX}

 
  \abstract
   {Analysing the large-scale structure (LSS) in the Universe with galaxy surveys demands accurate structure formation models. Such models should ideally be fast and have a clear theoretical framework in order to rapidly scan a variety of cosmological parameter spaces without requiring large training data sets.}
   {This study aims to extend Lagrangian perturbation theory (LPT), including viscosity and vorticity, to reproduce the cosmic evolution from dark matter $N$-body calculations at the field level.}
   {We extend LPT to a Eulerian framework, which we dub eALPT. { An ultraviolet regularisation through the spherical collapse model provided by Augmented LPT turns out to be crucial at low redshifts.} This iterative method enables modelling of the stress tensor and introduces vorticity. 
   The eALPT model has two free parameters apart from the choice of cosmology, redshift snapshots, cosmic volume, and the number of particles.}
   {We find that compared to $N$-body solvers, the cross-correlation of the dark matter distribution increases at $k = 1\,h\,{\rm Mpc}^{-1}$  and $z=0$ from $\sim55\%$ with the Zel'dovich approximation ($\sim70\%$ with ALPT), 
to $\sim$95\% with the three-timestep eALPT,  and the power spectra show percentage accuracy up to $k\simeq$ 0.3 $h\,{\rm Mpc}^{-1}$.}
{}
   \keywords{cosmology: – theory - large-scale structure of Universe - dark matter; methods: analytical}

   \maketitle
%

\section{Introduction}

The cosmic web is the large-scale structure (LSS) pattern that emerges as a manifestation of the action of gravity in an expanding background universe. It is shaped according to the cosmological information content set in the initial conditions of cosmic times. 

The scientific community is putting special effort into mapping the three-dimensional distribution of matter in the Universe through galaxy surveys such as DESI  \cite[][]{DESI}, EUCLID \cite[][]{Euclid},  J-PAS \citep{2014arXiv1403.5237B}, and the Nancy Grace Roman Space telescope \cite[][]{2015arXiv150303757S}.  The clustering of the LSS yields powerful constraints on the standard cosmological model, such as the nature of dark energy \citep[e.g.][]{2016arXiv161100036D},  primordial non-Gaussianities \citep[see, e.g.,][]{2019BAAS...51c.107M}, and neutrino masses \citep[see, e.g.,][]{2019JCAP...11..034C}.

Since the observable Universe is unique, one needs mock catalogues of it in order to perform robust analyses. Such catalogues permit the computation of covariance matrices and the study of systematics in the observational data. 
However, the computation of such a mock catalogue is usually costly, and only a few can be done \citep[see, e.g.,][]{2012MNRAS.426.2046A,2015MNRAS.448.2987F,2019MNRAS.487...48C}.
In this context, bias mapping techniques at the field level have emerged as a solution to save enormous computational resources while maintaining high accuracy. Such examples include \texttt{PATCHY} \citep[][]{2014MNRAS.439L..21K,2015MNRAS.450.1836K}, applied to the BOSS data \citep[][]{2016MNRAS.456.4156K}, and EZmocks \citep[][]{2015MNRAS.446.2621C}, which was applied to the eBOSS data \citep[][]{2021MNRAS.503.1149Z}.
More recently, the \texttt{BAM} code has been designed to learn the complex bias relation from reference simulations \citep[][]{2019MNRAS.483L..58B,2020MNRAS.491.2565B,10.1093/mnras/stac671}.
These techniques can also accurately map the Lyman-$\alpha$ forest \citep[see, e.g.,][]{2021ApJ...921...66S,2022ApJ...927..230S}. 
All the methods mentioned above require a dark matter field defined on a mesh. The accuracy of that matter distribution determines the precision of the mock catalogues.

A series of ideas have been implemented to accelerate particle-mesh based $N$-body solvers \citep[see][]{2013JCAP...06..036T,2016MNRAS.463.2273F}.
Despite these developments, $N$-body codes are costly when aiming to mass-produce mock catalogues covering large cosmic volumes. Therefore, approximate gravity solvers are still commonly used. While \texttt{EZmocks} relies on the Zel'dovich approximation \citep[][]{1970A&A.....5...84Z}, \texttt{PATCHY} and \texttt{BAM} rely on ALPT, including tidal field corrections \citep[][]{2013MNRAS.435L..78K}.
It has been shown that such methods can, to a great extent, correct the bias introduced by approximate gravity solvers in the nonlinear and non-local bias description \citep[see the application of \texttt{BAM} to galaxy catalogues,][]{2022arXiv221110640B}.

Machine learning techniques applied to large training data sets (of thousands of $N$-body simulations for a given cosmology) have emerged as an alternative to approximate gravity solvers \citep{2019PNAS..11613825H,2021PNAS..11820324D,2022arXiv220604594J}. (For a review of numerical methods in LSS modelling, see \citet[][]{2022LRCA....8....1A}.)
This work presents a novel approach to modelling the cosmic web by extending LPT approaches to an Eulerian framework. In the second section, we briefly review the theoretical background. In third section, we present the methodology, emphasising the ultraviolette regularisation, the Eulerian extension of LPT and the vorticity corrections. Section four discusses the numerical results. Finally, in section five, we present our conclusions.

\section{Theoretical background}

The Universe is considered a closed Hamiltonian system where energy is conserved. The dark matter content of the Universe can be described by a distribution function $f(\vec{r},\vec{v},t)$, with position $\vec{r}$ and velocity $\vec{v}$, such that the probability of finding a dark matter particle in the phase-space volume ${\rm d}\vec{r}{\rm d}\vec{v}$ centred on $\vec{r},\vec{v}$ at time $t$ is given by $f(\vec{r},\vec{v},t)\,{\rm d}\vec{r}{\rm d}\vec{v}$ \cite[see, e.g.,][]{2010gfe..book.....M}.
The total number of particles is then given by integrating over the whole phase-space volume: $N=\int\int {\rm d}\vec{r}{\rm d}\vec{v}\,f(\vec{r},\vec{v},t)$.
Given the functional form of the distribution function, total and continuous changes to the phase-space of the system of particles can be expressed with product and chain derivative rules:
$\frac{\rm d}{{\rm d}t}f=\frac{\partial}{{\partial}t}f+\frac{\partial\vec{r}}{{\partial}t}\frac{\partial}{{\partial}\vec{r}}f+\frac{\partial\vec{v}}{{\partial}t}\frac{\partial}{{\partial}\vec{v}}f=\frac{\partial}{{\partial}t}f+\vec{v}\frac{\partial}{{\partial}\vec{r}}f+{\vec{g}}\frac{\partial}{{\partial}\vec{v}}f$, with $\vec{g}$ being the gravity-induced acceleration.
Due to probability conservation, the generalised continuity equation in phase-space must be fulfilled (Liouville theorem):
$\frac{\partial f}{{\partial}t}+\nabla\cdot\vec{j}=\frac{\partial f}{{\partial}t}+\frac{\partial ( f\dot{\vec{q}})}{\partial \vec{q}}+\frac{\partial ( f\dot{\vec{v}})}{\partial \vec{v}}=0$. 
Hence, 
$\frac{\partial f}{{\partial}t}+\frac{\partial ( f\dot{\vec{q}})}{\partial \vec{q}}+\frac{\partial ( f\dot{\vec{v}})}{\partial \vec{v}}=
\frac{\partial f}{{\partial}t}+ f\frac{\partial \dot{\vec{q}}}{\partial \vec{q}}+\dot{\vec{q}}\frac{\partial  f}{\partial \vec{q}}+ f\frac{\partial \dot{\vec{v}}}{\partial \vec{v}}+\dot{\vec{v}}\frac{\partial  f}{\partial \vec{v}}=0$.
Consequently, to get the Vlasov or collisionless Boltzmann equation, the sum of terms multiplied by $f$ must vanish:
$f\left(\frac{\partial \dot{\vec{q}}}{\partial \vec{q}}+\frac{\partial \dot{\vec{p}}}{\partial \vec{p}}\right)= f\left(\frac{\partial^2 {\cal H}}{\partial \vec{q}\partial \vec{p}}-\frac{\partial^2 {\cal H}}{\partial \vec{p}\partial \vec{q}}\right)=0$, 
which is fulfilled by inserting Hamiltonian equations of motion, yielding:
\begin{equation}
\frac{\partial}{{\partial}t}f+\frac{\partial\vec{r}}{{\partial}t}\frac{\partial}{{\partial}\vec{r}}f+\frac{\partial\vec{v}}{{\partial}t}\frac{\partial}{{\partial}\vec{v}}f=\frac{\partial}{{\partial}t}f+\vec{v}\frac{\partial}{{\partial}\vec{r}}f+{\vec{g}}\frac{\partial}{{\partial}\vec{v}}f=0\,.
\end{equation}

Building moments of the Boltzmann equation, one gets the following: 
 From the zeroth moment, the continuity equation 
 multiplying Boltzmann's equation with the mass of the particles $m=m(\vec{v})^0$ and integrating over ${\rm d}\vec{v}$: $\frac{\partial}{{\partial}t}\overbrace{\int{\rm d}^3{v}\,f\,m}^{\rho}+\frac{\partial}{{\partial}\vec{r}}\overbrace{\int{\rm d}^3{v}\,f\,m\vec{v}}^{\rho\vec{u}}=0$, with $\rho$ being the density and $\vec{u}$  the velocity field. Hence, 
 \begin{equation}
\frac{\partial}{{\partial}t}\rho+\nabla\cdot(\rho\vec{u})=\frac{\partial}{{\partial}t}\rho+\rho\nabla\cdot\vec{u}+\vec{u}\cdot\nabla\rho=0\,.
\label{eq:continuity}
\end{equation}

 From the first moment, the Euler equation
 multiplying Boltzmann's equation with $m=m(\vec{v})^1$ and integrating over ${\rm d}\vec{v}$:
\begin{equation}
\frac{\partial}{{\partial}t}\rho\vec{u}+\frac{\partial}{{\partial}\vec{r}}(\rho\vec{u}\vec{u}+\rho\langle\vec{w}\vec{w}\rangle)=\rho\vec{g}\,,
\end{equation}
with $\vec{w}\equiv\vec{u}-\vec{v}$ being the random velocity.

The process of virialisation is expressed by a stress tensor $\tens{T}$ term: $\langle\vec{w}\vec{w}\rangle=-P\mathbf{1}$$+\tens{T}$.
The pressure term $-P\mathbf{1}$ can be neglected for collisionless dark matter.
From the combination of the Euler and continuity equations, we get:
$\rho\left(\frac{\partial}{{\partial}t}\vec{u}+\left(\vec{u}\cdot\nabla\right)\vec{u}\right)=\nabla\cdot\tens{T}+\rho\vec{g}$. Introducing co-moving coordinates $\vec{x}$: $\vec{r}=a(t)\vec{x}$, with the scale factor $a$ encoding the expansion of the Universe and conformal time $\tau$ determined by ${\rm d}t=a\,{\rm d}\tau$, we can rewrite the previous equation in terms of peculiar motions $\vec{v}\equiv a\dot{\vec{x}}$ instead of proper velocity $\vec{u}=\dot{\vec{r}}={\dot a}(t)\vec{x}+\vec{v}$, yielding:
\begin{equation}
    \frac{\partial}{\partial \tau }\vec{v}+\vec{v}\cdot\nabla\vec{v}=\frac{1}{a\rho}\nabla\cdot\tens{T}-\nabla\tilde{\Phi}-{a H}\vec{v}\,,
    \label{eq:euler}
\end{equation}
using  the Poisson equation $\nabla^2\tilde{\Phi}=4\pi G\,\bar{\rho}\delta a^2$ with density contrast $\delta\equiv\rho/\bar{\rho}-1$ and average density $\bar{\rho}\equiv\langle\rho\rangle$.
    
One usually assumes curl-free velocity fields, neglecting the stress tensor $\tens{T}$. Based on this, one can perform a Eulerian perturbative expansion of both the continuity and the Euler equations around the density contrast $\delta(\vec{x},\tau)=\sum_{n=1}^\infty\delta^{(n)}(\vec{x},\tau)$ and the divergence of the peculiar velocity field $\theta(\vec{x},\tau)\equiv\nabla\cdot\vec{v}=\sum_{n=1}^\infty\theta^{(n)}(\vec{x},\tau)$ \citep[see ][and references therein]{2002PhR...367....1B}. Alternatively, one can consider Eq.~\ref{eq:euler} and make an expansion in Lagrangian coordinates considering the total derivative $\frac{{\rm d}}{{\rm d}\tau}\vec{v}=\frac{\partial}{\partial \tau }\vec{v}+\vec{v}\cdot\nabla\vec{v}$, yielding Lagrangian perturbation theory (LPT) solutions \citep[][]{10.1093/mnras/264.2.375,1995A&A...296..575B,1995MNRAS.276..115C}. For $n$-th order LPT, see \citet{2021JCAP...04..033S}.

The stress tensor is commonly  represented by the linear elastic model $\nabla\cdot\tens{T}\sim\mu\nabla^2\vec{v}+\frac{
    1}{3}\mu\nabla\nabla\cdot\vec{v}+\eta\nabla(\nabla\cdot\vec{v})$ with viscosity parameters $\mu$ and $\eta$ \citep[see][]{2002PhR...367....1B}, which for irrotational fields (when the velocity is the gradient of a potential field) simplifies to the adhesion model  $\nabla\cdot\tens{T}\sim\mu'\nabla^2\vec{v}$ with a single viscosity parameter $\mu'$ \citep[see][]{RevModPhys.61.185}.
However, the generation of vorticity ($\nabla\times\vec{v}$)  has been studied in simulations as an important component when going to the non-linear regime \citep[see, e.g.,][]{2009PhRvD..80d3504P,2018JCAP...09..006J}.

As an alternative to expensive $N$-body simulations, effective field theories have emerged in LSS,  including a modelling of the stress tensor to compute summary statistics \citep[see, e.g.,][]{2012JHEP...09..082C,2012JCAP...07..051B,2013JCAP...08..037P,2014JCAP...05..022P,2014JCAP...03..006M,2015JCAP...10..039A,2015PhRvD..92l3007B,2016JCAP...03..017B,2016JCAP...05..027F}. (See also other perturbative  
  \citep[][]{2011JCAP...04..032M,2012JCAP...01..019P,Rampf_2012,2017PhRvD..95f3527C} and non-perturbative approaches  \citep[][]{2005A&A...438..443B} to modelling the curl.)

In this work, we propose modelling the viscosity induced by a stress tensor by iteratively applying LPT within a Eulerian framework with an ultraviolet 
regularisation given by the spherical collapse model. 
With this, the gravitational potentials become increasingly deeper and change their shape. In this way, the vorticity of the displacement field emerges naturally. 

\section{Method}

The method is based on a small-scale or ultraviolette regularisation obtained with Augmented Lagrangian perturbation theory, which is applied iteratively within a Eulerian framework.
Optionally, one may introduce vorticity corrections, as we discuss below.

We start by considering the Lagrangian $\vec{q}$ to Eulerian $\vec{x}$ coordinate single-step mapping through a displacement $\vec{\Psi}$ (with $\vec{v}=\frac{\rm d}{{\rm d}\tau}\,\vec{\Psi}$): $\vec{x}=\vec{q}+\vec{\Psi}(\vec{q})$.

\subsection{Regularisation of the small scales}

The displacement is obtained according to augmented LPT (ALPT), which separates the total displacement $\vec{\Psi}$ into a long-range $\vec{\Psi}_{\rm L}$ and a short-range component $\vec{\Psi}_{\rm S}$ \citep[see ][]{2013MNRAS.435L..78K}:
$\vec{\Psi}(\vec{q},z)=\vec{\Psi}_{\rm L}(\vec{q},z)+\vec{\Psi}_{\rm S}(\vec{q},z)$. 

The long-range displacement field is obtained from the convolution of a Gaussian kernel with an LPT solution:
$\vec{\Psi}_{\rm L}(\vec{q},z)=\mathcal{K}(\vec{q},z,r_{\rm s})\circ\vec{\Psi}_{\rm LPT}(\vec{q},z)$. We restricted the present  study to second-order LPT: $\vec{\Psi}_{\rm 2LPT}(\vq,z)=-D(z)\nabla_{\vq}\Phi^{(1)}(\vq)+D^{(2)}(z)\nabla_{\vq}\Phi^{(2)}(\vq)$, where $D(z)$ is the growth factor \citep[see, e.g.,][]{1977MNRAS.179..351H},
and $D^{(2)}(z)\simeq -\frac{3}{7}\,\Omega_{\rm m}^{-1/143}\left(D(z)\right)^{2}$ \citep[][]{1995A&A...296..575B}. The normalised potentials $\Phi^{i}(\vq)$ are the solutions of the Poisson equations $\nabla_{\vq}^{2}\Phi^{(i)}=\delta^{(i)}$, where $i=1$ is the linear primordial density field used as the initial conditions, and $i=2$ is determined by  $\delta^{(2)}=\sum_{i,j<i}\left( \partial_{ii}\Phi^{(1)}\partial_{jj}\Phi^{(1)}-(\partial_{ij}\Phi^{(1)})^{2}\right)$.

The short-range displacement is given by $\vec{\Psi}_{\rm S}(\vq,z)=\left(1-\mathcal{K}(\vq, r_{s})\right)\circ \vec{\Psi}_{\rm SC}(\vq,z)$, where the displacement $\vec{\Psi}_{\rm SC}(\vq,z)$ is derived within the spherical collapse (SC) approximation \cite[see][]{1994ApJ...427...51B},
$\psi_{\rm SC}(\vq,z)\equiv\nabla \cdot \vec{\Psi}_{\rm SC}(\vq,z)$, where $\psi_{\rm SC}(\vq,z)$ is the solution to the Poisson like equation \citep[see, e.g.,][]{2006MNRAS.365..939M,2013MNRAS.428..141N} $\nabla^{2}\psi_{\rm SC}(\vq,z)=3\left(\left[1-\frac{2}{3}D(z)\delta^{(1)}(\vq)\right]^{1/2}-1\right)$.

The short-range force has two regimes, one given by the spherical collapse model in the mild-density regime and one given by the perfect collapse to a point ($0=\nabla_{\vec{q}}\cdot\vec{x}=\nabla_{\vec{q}}\cdot(\vec{q}+\vec{\Psi})=3+\nabla_{\vec{q}}\cdot\vec{\Psi}$) in the high-density regime when the EPT approach yields imaginary solutions. 
The combination of both long- and short-range forces obtained through a kernel with a transition scale parameter $r_{\rm s}$) smears out the perfect collapse, partially emulating the result from quasi-virialisation. A second convolution with $r'_{\rm s}$ is applied to the primordial density field to determine the collapsing regions avoiding voids-in-clouds and is restricted to the high-density regime \citep[][]{2004MNRAS.350..517S,2016MNRAS.455L..11N}. Hence, the model has, so far, two free parameters, $r_{\rm s}$ and $r'_{\rm s}$, which are adjusted to improve the clustering towards small scales.  

\subsection{Eulerian augmented Lagrangian perturbation theory}

In this section, we generalise the approach to arbitrary Eulerian coordinates, substituting $\vec{q}$ by $\vec{x}_l$ for timestep $l$ so that $\vec{x}_{l+1}=\vec{x}_l+\vec{\Psi}(\vec{x}_l,z_{l+1})$ \citep[see also][]{2012MNRAS.425.2443K}. 
The initial dark matter particle positions in LPT are regularly distributed on a mesh ($\vec{q}$). In Eulerian-ALPT, however, the initial positions are given by the final positions of the previous step ($\vec{x}_{l}$).
After each step, we make a mass assignment of the dark matter particles onto a mesh to get the density contrast at each snapshot $\delta(\vec{x}_{l+1},z_{l+1})$. This field is used instead of the Gaussian field to compute the displacement field for the next step.

Due to mass conservation, we can write $\rho(\vec{x}_{l+1}){\rm d}\vec{x}_{l+1}=\rho(\vec{x}_{l}){\rm d}\vec{x}_{l}$.
Hence, $\frac{\rho(\vec{x}_{l+1})}{\rho(\vec{x}_{l})} =\big|\frac{{\rm d}\vec{x}_{l+1}}{{\rm d}\vec{x}_{l}}\big|^{-1}=\big|1+\nabla_{\vec{x}_{l}}\cdot\vec{\Psi}(\vec{x}_{l})\big|^{-1}$, where $\vec{x}_l=\vec{x}_{l-1}+\vec{\Psi}(\vec{x}_{l-1},z_{l})$ is not a homogeneous distribution after the first step. Nonetheless, we can perturbatively expand $\rho(\vec{x}_{l+1})\simeq\rho(\vec{x}_{l})(1+\delta(\vec{x}_{l+1}))$, which enables us to apply LPT within a Eulerian framework.
This implies that one should, in principle, use small step sizes between redshift calculations.  Only the first (A)LPT step is conveniently selected to be exceptionally long, as the initial conditions can be considered homogenous for this case.

If we consider two subsequent steps, we can write
$\vec{x}_{l+1}=\vec{x}_{l-1}+\vec{\Psi}(\vec{x}_{l-1},z_l)+\vec{\Psi}(\vec{x}_l,z_{l+1})$.
The total displacement to redshift $z_{l+1}$ is then given by $\vec{\Psi}(\vec{q},z_{l+1})\equiv\vec{x_{l+1}}-\vec{q}=\vec{\Psi}(\vec{q},z_{l-1})+\vec{\Psi}(\vec{x}_{l-1},z_l)+\vec{\Psi}(\vec{x}_l,z_{l+1})$.
We assume curl-free fields in each step. Hence, $\nabla_{\vec
{x}_m}\times\vec{\Psi}(\vec{x}_m,z_{m+1})=0$ for $m=1,\dots,n$, with $n$ being the total number of redshift snapshots.
The shape of the gravitational potentials changes with redshift, $\Phi^{(i)}|_{z_{l+1}}\,\cancel{\propto}\,\Phi^{(i)}|_{z_l}$ for $i=1,2$.
Therefore,  $\nabla_{\vec
{q}}\times\vec{\Psi}(\vec{q},z_{l+1})\neq0$ for $l>0$, and $z_1$ is the first redshift snapshot. 
Hence, already after two steps, vorticity at Lagrangian coordinates emerges naturally. 
For this reason, it is wrong to assume that there are no curl sources after one step, even if one assumes curl-free fields initially.

\subsection{Vorticity corrections}

{ In particular, in each timestep we neglect the stress tensor term $\tens{T}$ and the divergence-free velocity component  in Eq.~\ref{eq:euler}.}
We can compensate for this vorticity leakage by adding in the subsequent step $l+1$ a fraction of the vorticity generated in the previous step $l$, which is computed as the difference between the total displacement to redshift $z_{l}$ and the corresponding curl-free component. The displacement field can be decomposed according to the Helmholtz theorem into an irrotational or longitudinal (curl-free) and a solenoidal or transversal (divergence-free) component: $\vec{\Psi}=\vec{\Psi}_{\nabla\times\vec{\Psi}=0}+\vec{\Psi}_{\nabla\cdot\vec{\Psi}=0}$.

\begin{figure}
\includegraphics[width=0.45\textwidth]{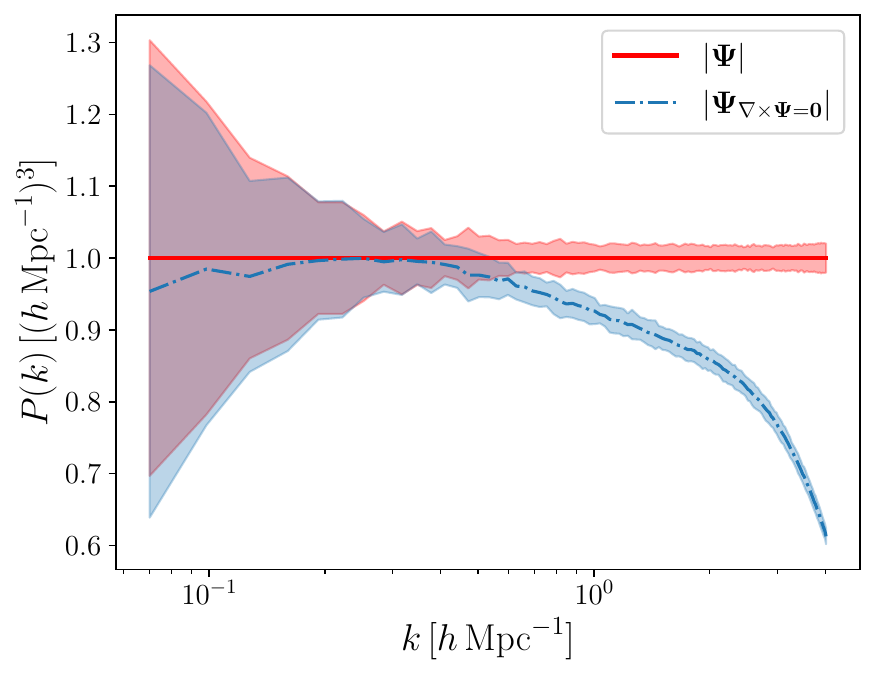}
\put(-90,50){$z=0$}
\caption{\label{fig:helmholtz}  Power spectra of the displacement fields from the Helmoltz decomposition for the FastPM simulations at $z=0$ (mean and standard deviation) comparing the total component $|\vec{\Psi}|$ to the longitudinal one $|\vec{\Psi}_{\nabla\times\vec{\Psi}=0}|$. The 1-sigma contours have been computed from 25 FastPM simulations.}
\end{figure}

To get the solenoidal component, we exploit the fact that the divergence of the curl of a vector field vanishes:
\begin{equation}
\vec{\Psi}_{\nabla\cdot\vec{\Psi}=0}(\vec{q},z_{l})\equiv \left(1-\nabla\nabla^{-2}\nabla\right)\cdot\vec{\Psi}(\vec{q},z_{l})\,, \end{equation}
where $\vec{\Psi}_{\nabla\times\vec{\Psi}=0}\equiv\nabla\nabla^{-2}\nabla\cdot\vec{\Psi}(\vec{q},z_{l})$.

{From the third step on ($l>1$), we naturally get a transversal component that is not properly evolved in each step according to  Eq.~\ref{eq:euler}. A potential way of improving this could be achieved by introducing a transfer function ${\cal T}(\vec{q},z_l)$ in the following way \cite[see also][]{2020MNRAS.498.2663T}:
\begin{equation}
\vec{x}_{l+1}=\vec{x}_l+\vec{\Psi}(\vec{x}_l,z_{l+1})+{\cal T}(\vec{q},z_l) \circ  \vec{\Psi}_{\nabla\cdot\vec{\Psi}=0}(\vec{x}_l(\vec{q}),z_{l})\,.
\end{equation}
We consider a reduced version of the transfer function to a single scalar value, a third free 
parameter $\alpha$, in order to modulate the amplitude of the transversal component:
}
\begin{equation}
\vec{x}_{l+1}=\vec{x}_l+\vec{\Psi}(\vec{x}_l,z_{l+1})+\alpha \vec{\Psi}_{\nabla\cdot\vec{\Psi}=0}(\vec{x}_l(\vec{q}),z_{l})\,.
\end{equation}
We explore the cases with and without $\alpha$ boost contributions in the next section.

\begin{figure*}
 \centering
\begin{minipage}{.6\textwidth}
\hspace{0.5cm}
\begin{tabular}{c}
\includegraphics[width=0.8\textwidth]{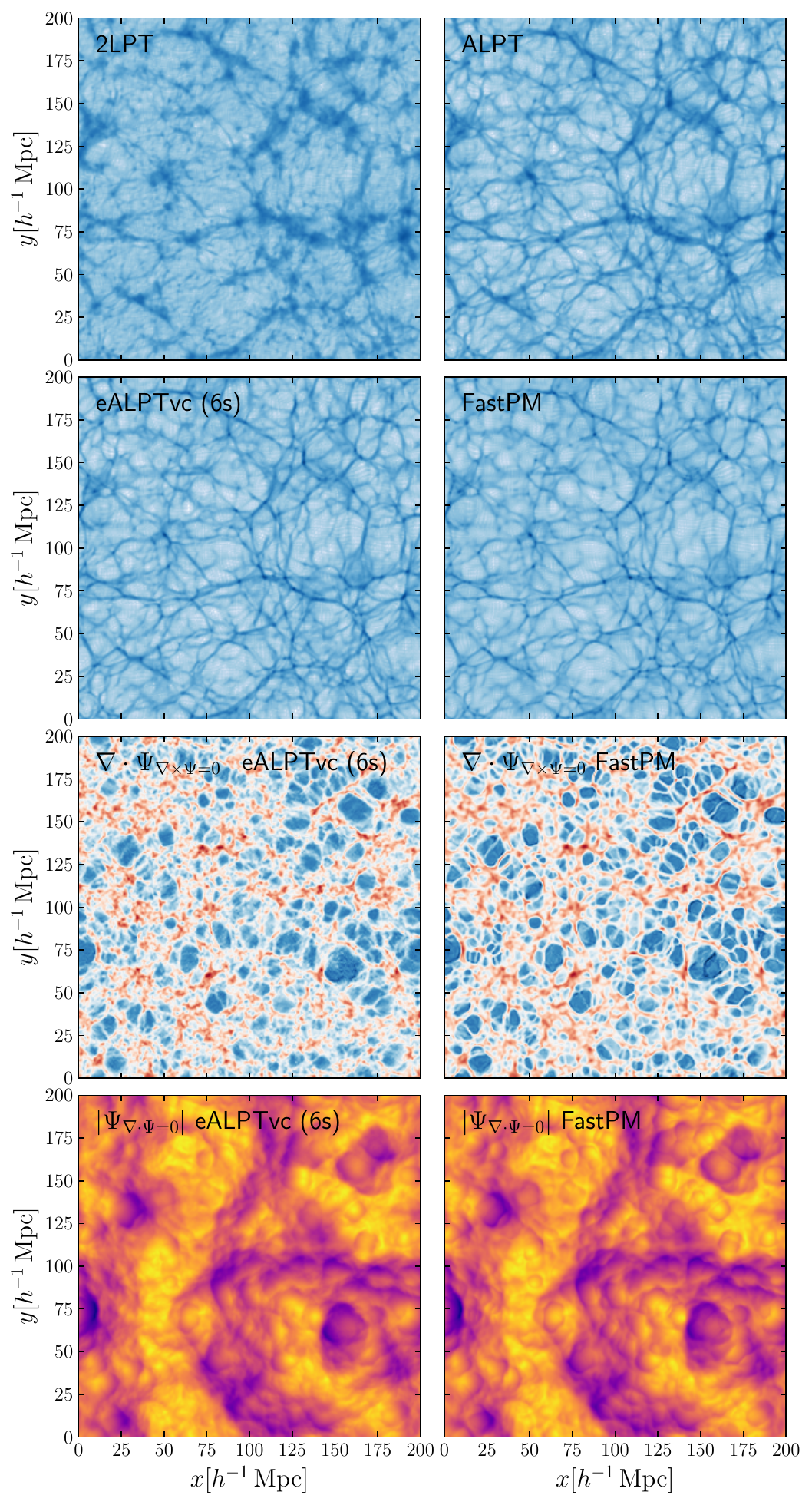}
\end{tabular}
\vspace{-0.5cm}
\caption{\label{fig:slices} Cosmic density and displacement field components as obtained with different methods. Upper panels: Log-density  slices  at redshift $z=0$ ($\log(1+\delta)$) corresponding to different gravity solvers 
through the three-dimensional simulation cubical box of 200 $h^{-1}$ Mpc side length with a mesh of 256$^3$ cells.  Lower panels: Same as the upper panels but for the divergence of the irrotational  ($\nabla_{\vec{q}}\cdot\vec{\Psi_{\nabla\times\vec{\Psi}=0}}(\vec{q})$) 
and the absolute value of the solenoidal component of the displacement field,  $\big|\vec{\Psi}_{\nabla\cdot\vec{\Psi}=0}(\vec{q})\big|$ (with power spectra within percentage agreement up to $k\simeq$ 0.4 $h\,{\rm Mpc}^{-1}$).
The resulting maps have been saturated to the same colour scale after averaging over five slices corresponding to one simulation.}
\end{minipage} 
\hspace{0.3cm}
\begin{minipage}{.3\textwidth} 
\includegraphics[width=\textwidth]{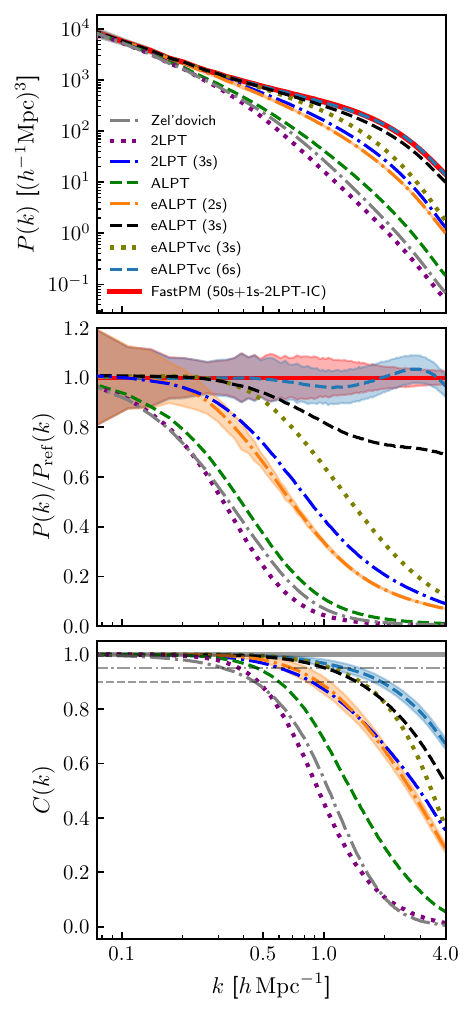}
  \put(-40,315){$z=0$}
  \caption{\label{fig:pkandc}Power and cross power spectra at redshift $z=0$ (averaging in $\Delta k$-bins for fields A and B: $C(k)=\langle\hat{\delta}_{\rm A}(\vec{k})\overline{\hat{\delta}_{\rm B}(\vec{k})}\rangle_{\Delta k}/\sqrt{P_{\rm A}(k)P_{\rm B}(k)}$ ) for the various approximate gravity solvers  compared to fast particle-mesh based $N$-body calculations (FastPM). The Zel'dovich, 2LPT, and ALPT cases require only a one-step calculation. For the rest, the number of steps is indicated in parentheses. FastPM needs the first step for the ICs relying on 2LPT and then a minimal number of iterations compared to standard $N$-body simulations, which require iterations on the order of $10^3$. 
 Also shown are the 1-sigma contours from 25 fields for FastPM and the eALPT-limiting cases with the lowest and highest cross-correlations, two (w/o vc) and six (with vc) steps, respectively. The  
 eALPT (3s) and the eALPTvc (6s) cases yield power spectra with precision within a few per cent up to $k\simeq$ 0.3  and 4 $h\,{\rm Mpc}^{-1}$ (the Nyquist frequency) as well as correlations of 95\% and 98\% at $k\simeq1\,h\,{\rm Mpc}^{-1}$, respectively.}
\end{minipage}
\end{figure*}

\begin{figure}
\includegraphics{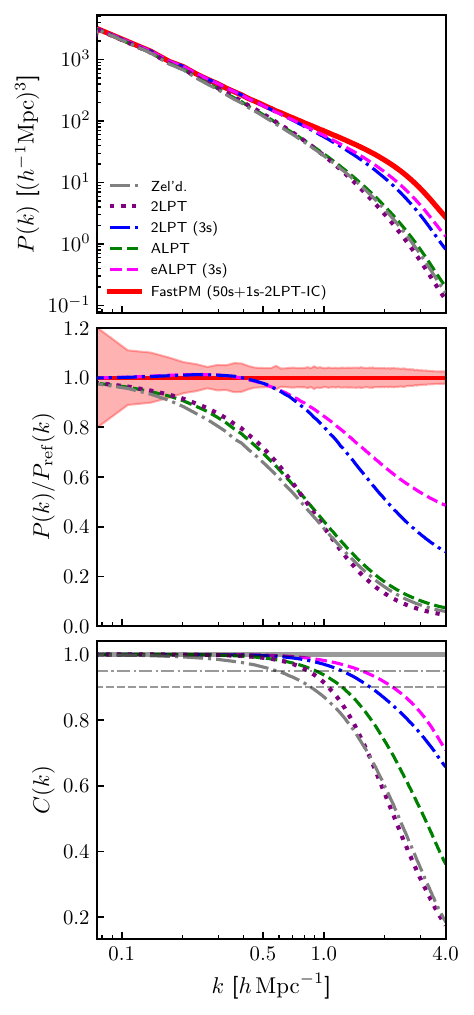}
\put(-50,45){$z=1$}
\caption{\footnotesize\label{fig:pkandcz1}Power and cross power spectra  at redshift $z=1$  analogously to Fig.~\ref{fig:pkandc}. }
\end{figure}

    \section{Numerical results}

We consider in this study cubical boxes with a side length of 200 $h^{-1}$ Mpc and 256$^3$ particles using Lambda-CDM cosmology with an initial power spectrum obtained from CLASS \citep{2011arXiv1104.2932L,2011JCAP...07..034B} with the cosmological parameters $\{\Omega_M=0.307, \Omega_\Lambda=0.693,\Omega_b=0.048, \sigma_8=0.823, w=-1, n_s=0.95\}$ and a Hubble constant ($H_0 = 100 \,h$   km s$^{-1}$ Mpc$^{-1}$) given by $h$ = 0.68.
We constructed two reference sets of 25 $N$-body calculations, each at $z=0$ and $z=1$,  relying on FastPM \citep[][]{2016MNRAS.463.2273F} with the same volume and number of particles, the force resolution $\sim 0.39 \, h^{-1}\,\rm{Mpc}$, and 50 timesteps. Initial conditions were generated via second-order LPT at $z=99$.

Using the same Gaussian initial density field, we ran eALPT with ($\alpha>0$) and without  ($\alpha=0$) vorticity corrections (vc), which we dubbed eALPT and eALPTvc, respectively. In particular, we studied the cases of eALPT with two, three, and five steps and eALPTvc with three, five, and six steps (see Figs.~\ref{fig:slices}-\ref{fig:pkandcz1}).

{Figure~\ref{fig:helmholtz} shows the Helmholtz decomposition for the FastPM displacement fields at $z=0$.  We find that the difference in power between the full and the longitudinal components is below 10\% for $k<1\,h\,\text{Mpc}^{-1}$. Hence, focusing on the range of scales below $k\sim1\,h\,\text{Mpc}^{-1}$,  the alpha vorticity boost parameter cannot be larger than 10\% in each timestep and should decrease further with an increasing number of timesteps.}

The size of each eALPT (or eALPTvc) timestep is computed based on the difference between the linear growth factors from subsequent steps: 
\begin{equation}
    \Delta D(z_{l+1},z_l)\equiv D(z_{l+1})-D(z_l)\,.
\label{eq:timestep}
\end{equation}
    
For a given cosmology, we determine the corresponding redshift: $\Delta z(z_{l+1},z_l)=D^{-1}(\Delta D(z_{l+1},z_l))$, inverting the $D=D(z)$ relation through an interpolation based on a previously computed table. We then use $\Delta z$ for $D(\Delta z)$ and  $D^{(2)}(\Delta z)$ to compute the displacement field to go to redshift $z_{l+1}$ starting at redshift $z_l$ in Eulerian coordinates. 

Because of numerical uncertainties arising from the mass-assignment scheme and from using significant steps between Eulerian coordinates at subsequent redshifts, the power at large scales of the resulting density field does not correspond to the one expected from linear theory.
Hence, the step size has to be adjusted to get a precise final result. To this end, single-step ALPT calculations can be applied as a reference to determine excess or lack of power following \citet[][]{2021MNRAS.502.3456K} using Gaussian convolutions of the density field to determine the variance and thus the resulting bias: 
\begin{equation}
b(z_{l+1},z_l)=\sqrt{\frac{\langle ({\mathcal K}(r''_{\rm S})\circ\delta^{\rm eALPT}(\vec{x}_{l+1},z_{l+1}))^2\rangle}{\langle ({\mathcal K}(r''_{\rm S})\circ\delta^{\rm ALPT}(\vec{x}_{l+1},z_{l+1}))^2\rangle}}\,,
\label{eq:bias}
\end{equation} 
with $r''_{\rm S}\in[30,70] \,h^{-1}\,{\rm Mpc}$.
We note that the exact choice of $r''_{\rm S}$ is not critical, and results within percentage accuracy are obtained for different values in that range. The timestep size has to be accordingly corrected: $\Delta z(z_{l+1},z_l)=D^{-1}(\Delta D(z_{l+1},z_l)-(b(z_{l+1},z_l)-1))$. These corrections must be computed only once for a particular setup and can then be applied for different seed perturbations. 

We find that the resulting density fields are very stable, relying on either cloud-in-cell or triangular shaped cloud (TSC) mass-assignment schemes \citep[][]{1988csup.bookH} with and without tetrahedral tesselations (THT) \citep[][]{2012MNRAS.427...61A}. Finally, we used TSC with THT.

From our numerical tests, we draw following conclusions (see Figs.~\ref{fig:slices},  \ref{fig:pkandc} and \ref{fig:pkandcz1}):
\begin{enumerate}
    \item  The ideal transition scale parameter $r_{\rm s}$ between long- and short-range forces has a lower $r_{\rm s}>r_{\rm s-min}$ and an upper limit of  $r_{\rm s}<r_{\rm s-max}$.  While $r_{\rm s-min}$ has to be greater than zero to suppress shell-crossing in knots of the cosmic web, the accuracy in the tidal field tensor limits $r_{\rm s-max}$. In practice, we find $2<r_s<10$ $h^{-1}$ Mpc.

\item The scale $r'_{\rm s}$ is  restricted to the range $0\leq r'_{\rm s}\leq2\,{\rm d}L$, where ${\rm d}L$ is the cell side length. Neither larger smoothing scales nor additional convolutions have any impact on the results.

\item The fraction of additional vorticity is constrained to the range $\alpha\in[1,10]$ [\%] and requires cross-correlations with the reference simulation, as a different combination of parameters can yield similar power spectra. In particular, the vorticity boost can partially compensate for the ultraviolet regularisation (compare numerical values of $r_{\rm S}$ with and without $\alpha$ below). We find for the three-step case at $z=0$ that the vorticity boost very moderately improves the results with $\alpha=0$ up to $k\simeq$ 1.5 $h\,{\rm Mpc}^{-1}$in terms of cross-correlation and are already worse in terms of power spectrum for $k>$ 0.4 $h\,{\rm Mpc}^{-1}$. It is only by applying more timesteps that we can benefit from the vorticity boost. 

\item The three-step eALPT calculations neglecting the vorticity boost do not require any reference simulation and achieve percentage accuracy in the power spectrum up to $k\simeq$ 0.3 $h\,{\rm Mpc}^{-1}$ and $\sim$95 \% cross-correlation at $k=$ 1 $h\,{\rm Mpc}^{-1}$. 

\item The three-step 2LPT calculation considerably improves the clustering towards small scales compared to the one-step 2LPT, but it lacks an ultraviolet regularisation, which significantly limits its accuracy towards high $k$-values, especially at low redshifts (compare results from Figs.~\ref{fig:pkandc} and \ref{fig:pkandcz1}). 

\item The $z=1$ calculations show that the iterative application of LPT at subsequent times considerably improves the results compared to the one-step calculations. We also find that the ultraviolet regularisation becomes less critical. 

\item The optimal timestep choice consists of first making a long step with ALPT and significantly smaller subsequent steps. However, the timesteps need to be large enough to avoid numerical uncertainties arising from being too short (see Eq.~\ref{eq:timestep}). The last timestep can be adjusted to ensure a vanishing bias, that is, $b\to1$ in Eq.~\ref{eq:bias} (in our calculation, we demand a precision of better than 0.5\% in $b\simeq1$). This greatly simplifies the calibration process for each set of free parameters. 

\end{enumerate}
The parameters are chosen in order to obtain unbiased power spectra starting from the lowest modes and continuing towards the largest ones.

The concrete values for the eALPT runs are the following: For two steps, $\{r_{\rm s}=5,r'_{\rm s}=2,\alpha=0\}$; for three steps, $\{r_{\rm s}=6,r'_{\rm s}=2,\alpha=0\}$ and $\{r_{\rm s}=3.1,r'_{\rm s}=1.2,\alpha=7\}$; for five steps, $\{r_{\rm s}=7,r'_{\rm s}=2,\alpha=0\}$ and $\{r_{\rm s}=3,r'_{\rm s}=2,\alpha=3.5\}$; and for six steps, $\{r_{\rm s}=2.75,r'_{\rm s}=1.4,\alpha=3\}$. 
The numerical results demonstrate a higher accuracy when applying vorticity corrections but at the expense of depending on reference simulations, as the results need to be cross-checked with reference full-volume density fields. 
We also find that vorticity partially substitutes for the short-range component, as the transition scale parameter $r_{\rm S}$ becomes smaller for $\alpha>0$. 
{As expected from our previous discussion, the optimal $\alpha$ parameter remains below 10\%. In fact, the first time we can introduce $\alpha$ is in a third timestep calculation, since vorticity emerges only after two timesteps, and we found $\alpha=7\%$  and even lower values for more timesteps. We checked the power spectra of the solenoidal component for both the reference FastPM runs and the eALPTvc(6s) calculations, and we found an excellent agreement, within percentage accuracy, up to $k\sim0.5\,h\,\text{Mpc}^{-1}$. However, the deviations for that component can become larger than 10\% towards higher frequencies. }
We also checked that we obtain an equivalent level of accuracy with different cosmological parameters.

\section{Conclusions}

Accurate calculations of the cosmic web dark matter distribution can be made very quickly with effective models by iteratively applying LPT within a Eulerian
framework with an ultraviolet regularisation given by the
spherical collapse model, which we dub eALPT.   
We showed that within three steps we can already find cross-correlations at the level of $\sim$95\% at $k=1\,h\,{\rm Mpc}^{-1}$  and power spectra within a percentage accuracy up to $k\simeq0.3\,h\,{\rm Mpc}^{-1}$ as compared to $N$-body calculations. Moreover, this model is only about three to four times more expensive than setting up initial conditions for an $N$-body simulation.
We also explored a variation of the eALPT method by adding a small fraction of the vorticity in each subsequent timestep, finding some  improvement; however, it is at the expense of requiring calibration with a reference simulation.

We plan to use the method studied in this work to mass-produce lightcone dark matter fields covering the entire redshift range up to $z\sim 4$, populating them with bright galaxies, luminous red galaxies,  emission line galaxies, quasars, Lyman-$\alpha$ forests, and lensing maps in order to conduct a multi-tracer analysis from galaxy surveys.
The developments presented in this work have applications ranging from setting up initial conditions, producing mock catalogues for galaxy surveys, and performing Bayesian inference analysis to providing simulations for emulators. Nevertheless, further investigation needs to be done to explore the accuracy within this framework in modelling the peculiar velocity field and to study different resolutions and redshifts.

\begin{acknowledgements}
FSK, FS, ABA, and GF acknowledge  the Spanish Ministry of Economy and Competitiveness (MINECO) for financing the \texttt{Big Data of the Cosmic Web} project: PID2020-120612GB-I00 and the IAC for continuous support to the \texttt{Cosmology with LSS probes} project. FS thanks financial support from the University of Padova, ABA thanks support from the SEV-2015-0548 grant, and GF thanks support from the IJC2020-044343-I grant.
\end{acknowledgements}

%
%
\bibliographystyle{aa}
\bibliography{refs}  

\end{document}